\documentclass[doublecol]{epl2}
\usepackage{graphicx}
\usepackage{amsmath}
\usepackage{amssymb}
\usepackage[utf8]{inputenc}
\usepackage{comment}
\usepackage{color}
\usepackage{dcolumn}
\usepackage{bm}
\usepackage{hyperref}
\usepackage{soul}
\usepackage{footmisc}
\usepackage{braket}

\newcommand{\Rmnum}[1]{\expandafter\@slowromancap\romannumeral  #1@}

\usepackage{bbold}

\newcommand{\sy} \revision{\color{black}}
\graphicspath{{pics/}}

\title{Nonequilibrium Kondo-vs-RKKY Scenarios in Nanoclusters}
\shorttitle{Nonequilibrium Kondo-vs-RKKY Scenarios in Nanoclusters} 

\author{Simon Ydman\inst{1,2} \and Miroslav Hopjan\inst{1,2} \and Claudio Verdozzi\inst{1,2}}
\shortauthor{S. Ydman \etal}

\institute{                    
  \inst{1} Department of Physics, Lund University, PO Box 118, 221 00 Lund, Sweden\\
  \inst{2} European Theoretical Spectroscopy Facility (ETSF)
}

\abstract{
Ultrafast manipulations of magnetic phases are eliciting increasing attention from the scientific community, because potentially relevant to the understanding of nonequilibrium phase transitions and to novel technologies. Here, we focus on manipulations applied to magnetic impurities in metallic hosts. By considering small nanoring geometries, we show how currents can induce a dynamical switching between different types of exchange interactions in these systems. Our work thus opens a study window on nonequilibrium Doniach's magnetic phase diagrams, and time-dependent Kondo-vs-RKKY scenarios.}

\begin{document}
\maketitle
{\bf Introduction. -} Magnetic phenomena are known to man from antiquity \cite{Chinese},
with documented references to lodestone as early as nearly three millennia ago.
A proper notion of the basic nature of magnetism, though, is a much more recent outcome,  
made possible in the last century by the advent of quantum mechanics.
Then it became clear that it is the exchange interaction \cite{HeiseExch,DiracExch},
a produce of Coulomb force and symmetry properties of the many-particle wavefunction, to be responsible for the ordering of magnetic moments (e.g. electron spins) in a system. 

Nowadays, it is possible to create and employ ultrashort electromagnetic pulses to probe and control quantum mechanical systems and alter nature and strength of their many-body interactions, making feasible the study of dynamical phase transitions \cite{Jurcevic2017}. 
For magnetism, this means a direct manipulation of the exchange interaction, to attain different magnetic orders and be able to dynamically switch between them \cite{Kimel,Mentink,Batignani} for technological aims. 

A type of systems which is attractive in this context is localized magnetic impurities in metallic hosts: due to strong electronic correlations, they exhibit a fascinating behavior, depending on temperature, impurity density, and disorder \cite{Doniach77,Varma,Smith2000,Hewson}. For their magnetic properties, a key feature is the interplay of Kondo screening (due to the exchange coupling between conduction electrons and magnetic moments at the impurities) and Ruderman-Kittel-Kasuya-Yoshida (RKKY) exchange (the conduction electrons induce an indirect magnetic coupling among the moments at the impurities) \cite{Hewson,Fazekas}. This competition provides the rationale behind the so-called Doniach-Phase diagram \cite{Doniach77}, which shows how different magnetic phases relate to temperature and exchange interaction.

This subject has received great attention for bulk samples, in connection to heavy-fermions and mixed-valence compounds \cite{Hewson,Fazekas}. However, finite samples are also of high importance, in relation to nanotechnology \cite{Odom2000,Pastor,Tyagi,Haraldsen,Domanski,Eriksson,Rubio}. Finite-size effects were initially studied for isolated impurities in the so-called Kondo-box model \cite{KondoBox}, with the average conduction level spacing 
introducing a new energy scale. For dense impurities, such spacing was found to be a regulating factor to the interplay of Kondo and RKKY interactions \cite{Verdozzi2004}, leading to a Doniach-like magnetic phase diagram for nanoclusters \cite{Luo2005,Samuelsson2007,Potthoff1,Potthoff2}.

These studies on dense impurities focussed on systems in equilibrium; 
it could then be of interest to consider the dynamical regime. 
This possibility was recently pointed out for bulk samples \cite{Takasan17}; however, the case of finite systems, where a finite conduction level spacing plays a role, has remained unexplored so far.
This is the issue we wish to address here: Namely, {\it can we dynamically manipulate the interplay of Kondo and RKKY exchange interactions in finite systems with dense magnetic impurities?} 

Our answer is in the affirmative: By studying nanorings with dense magnetic impurities, and using magnetic fluxes to induce ring currents in the conduction levels, we show that i) it is possible to go across different magnetic configurations in time, and in fact ii) realise a non-equilibrium Doniach-like phase diagram. 
We also iii) show how such crossing is driven effectively via optimal control theory \cite{Gross}; furthermore, since in general the exchange interactions depends on cluster shape/size and impurity distribution, one can have a broad palette of magnetic switchings, with new potentialities
to use Kondo-vs-RKKY scenarios in practical applications.\\
\begin{figure}[tbh]
  \centering
\includegraphics[width=0.47\textwidth]{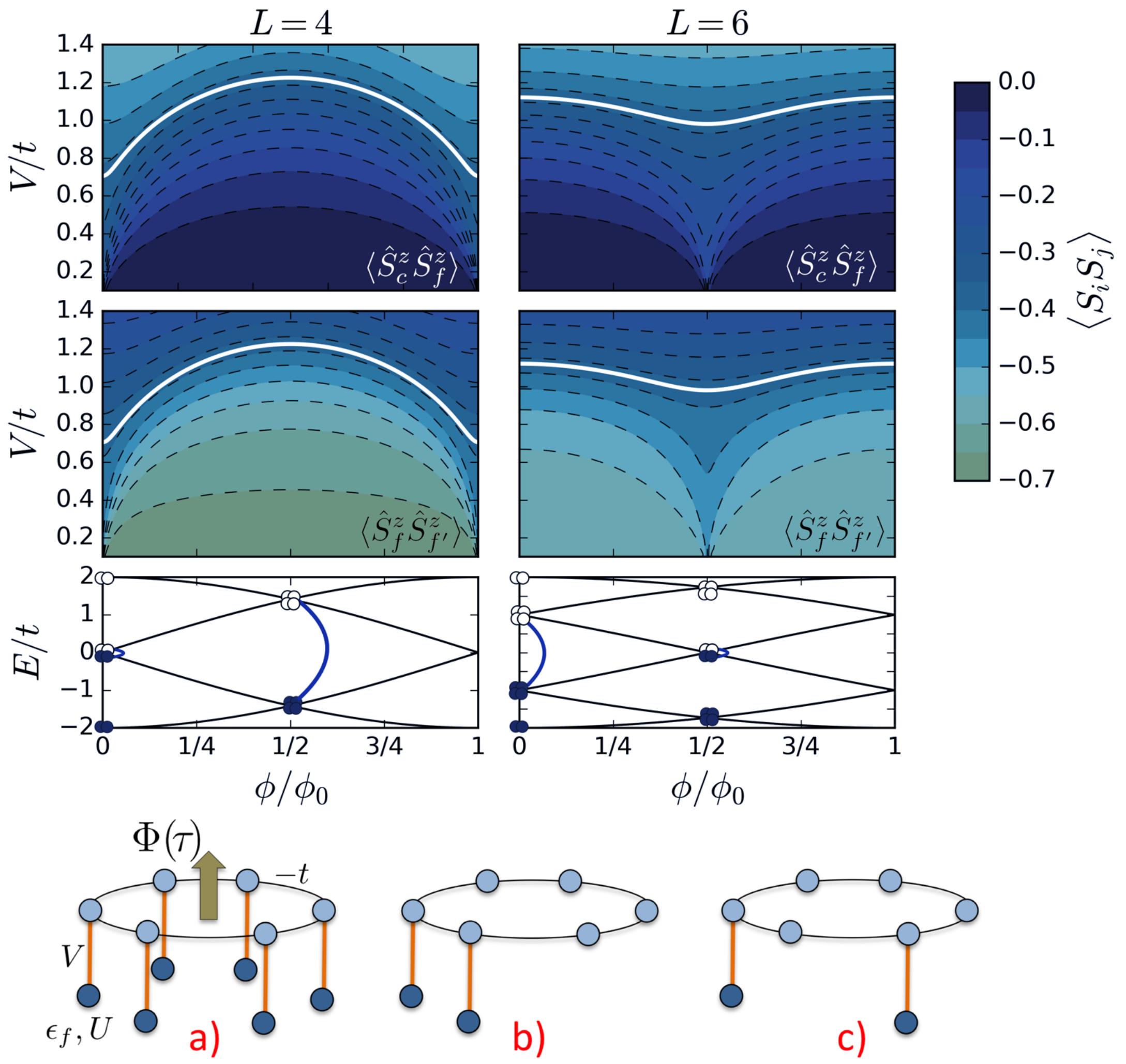}
  \caption{Kondo (top row panels) and RKKY (second row panels) correlations of 4-site (left) and 6-site (right) spin-compensated clusters as a function of the Peierls phase $\phi/\phi_{0}$. White
lines show the border between Kondo and RKKY regimes. \revision{\color{black}The scale for the strength of the correlations is reported in the vertical bar on the right, whilst the dashed lines in the diagrams are a guide to the eye.}
In the third row panels we show energy states for conduction electrons together
with level occupation when $\phi/\phi_{0}=0$ and $\phi/\phi_{0}=0.5$. \revision{\color{black} The thin lines represent
how the single particle energy level (four for $L=4$ and six for $L=6$) evolve as function of $\phi/\phi_{0}$. The blue
(white) circles represent occupied (empty) levels, and the blue arches show the energy gap at specific value of the phase } The bottom diagrams show the PAM cluster for $L=6$ in the a) dense- and b,c) dilute-impurity case. \revision{\color{black} In this case, light (dark) blue circles denote conduction (impurity) orbitals, whilst the black arches schematically denote the hopping terms between nearest-neighbour conduction orbitals, and the orange lines denote the hybridization terms between conduction and impurity orbitals at the same site. The symbols labeling these quantities are defined in the system's Hamiltonian, Eq.~(\ref{eq:hamiltonian}).}}
  \label{fig:kondo}
\end{figure} 
{\bf Model and method.-} We consider rings with $L=4,6$ sites, two orbitals/site 
(labeled $c/f$ for conduction/impurity levels), and with $2L$
electrons (half-filling).  Theoretically, ring topology is attractive since it admits periodic boundary conditions and steady-state currents in small samples. Further, nanorings with only few electrons \cite{Lorke2000,Sadowski,Rings} in magnetic fields \cite{Aharonov1959,Kasai2002} can be
created experimentally. 

Our rings are described by a Periodic Anderson Model (PAM) Hamiltonian (see Fig.~\ref{fig:kondo}a), with a (possibly dependent on time $\tau$) magnetic flux $\Phi$ perpendicular to the rings, included via Peierls phases in the kinetic energy term of the conduction electrons:
\begin{align}
&H(\tau)=-t\sum_{\langle i,j\rangle, \sigma}e^{i \phi_{ij}(\tau)}c_{i\sigma}^{\dagger} c_{j\sigma}^{}+ \epsilon_{f} \sum_{i,\sigma}f^{\dagger}_{i\sigma}f_{i\sigma}^{} \nonumber\\
&+U\sum_{i}f^{\dagger}_{i\uparrow}f^{}_{i\uparrow} f^{\dagger}_{i\downarrow}f^{}_{i\downarrow} 
+V \sum_{i,\sigma}( c_{i\sigma}^{\dagger}  f_{i\sigma}^{}+ h.c. ),
\label{eq:hamiltonian}
\end{align}
where $\langle ... \rangle$ denotes nearest-neighbour sites, $c_{i\sigma}^\dagger$ ($f_{i\sigma}^\dagger$) creates a conduction (impurity) electron of spin projection $\sigma$ at
site $i$, and $V$ is the hybridization between impurity and conduction level at the same site. 
$U$ and $\epsilon_{f}$ are onsite interaction and orbital energy at the impurity site, respectively.  We set $2\epsilon_f+U=0$ (the particle-hole symmetric Kondo regime),
with one electron/$f$-orbital, and $L$ electrons in the conduction orbitals (this
maps the PAM into a Kondo lattice model).
Also, $\phi_{ij}(\tau) = -\phi_{ji}(\tau) = \phi({\tau})=\Phi(\tau)/L$ with period
$\phi_{0}=2\pi/L$.
To determine the ground state and the non-equilibrium dynamics of the system, we use exact diagonalization based on the Lanczos algorithm \cite{Lancz1,Lancz3}.  
For a given operator $\hat{O}$, thermal averages at temperature $T$ are computed as
$\langle \hat{O} \rangle = \text{Tr}[\hat{O}\hat{\rho}]$, 
with $\hat{\rho}=e^{-\mathcal{H}/k_BT}/ \text{Tr}[e^{-\mathcal{H}/k_BT}]$,  
$\mathcal{H}= H(\tau=0)$.  The optimally controlled time-evolution is done at $T=0$, using the GRAPE algorithm \cite{Manches2011,Fouquieres2011}, and maximizing the fidelity function $\mathcal{F}(\phi) = |\braket{\psi_{tar} | \psi_{\tau_M}(\phi)}|$ with respect to the control parameter $\phi(\tau)$. \revision{\color{black}  $\mathcal{F}(\phi)$ gives an indication of the ``resistance'' of the system to switch from one phase to the other, and also quantifies the degree of success in obtaining the target phase.} In $ \mathcal{F}$, $\psi_{\tau_M}$ is the time-evolved state at final time $\tau_M$, $|\psi_{tar}\rangle$ is the chosen target state, and the Hamiltonian can be recast in generic
form as $H(\tau)=H_{0}+V_{oc}(\phi)$. $H_{0}$ can depend on time, but
optimization is only performed on $V_{oc}$.  In practice, time-discretization is used, with $\tau_M=N\Delta$ and $n=1,...N$. Then, $\phi(\tau)\rightarrow \phi(\tau_n)\equiv \phi_n $ and
$H(\tau)\rightarrow H(\tau_n,\phi_n)$, to be optimized for $(n-1)\Delta \le\tau\le n\Delta$. Given an initial state $|\psi_{i}\rangle$, we define the evolution operator via
$| \psi_{\tau_M}\rangle = \hat{U}(\tau_M,0)|\psi_{i}\rangle$. Discretizing, $\hat{U}(\tau_M,0)\!=\!e^{-iH(\tau_N)\Delta}e^{-iH(\tau_{N-1})\Delta}\dots e^{-iH(\tau_1)\Delta}$, so that 
\begin{align}\label{overl1}
\braket{\psi_{tar} | \psi_{\tau_M}}=
\langle\psi^{(n+1)}_{tar}|e^{-i[H(\tau_{n},\phi_n)]\Delta}|\psi^{(n-1)}_{init}\rangle. 
\end{align}
Here, $\langle\psi^{(n+1)}_{tar}|\equiv \langle\psi_{tar}|\hat{U}(\tau_M,\tau_{n+1})$ and $|\psi^{(n-1)}_{init}\rangle\equiv
\hat{U}(\tau_{n-1},0)|\psi_{i}\rangle$.
To obtain $\partial_{\phi_n} \mathcal{F}$, we need
\begin{align}
\!\!\!\!\partial_{\phi_{n}}\!\!\braket{\psi_{tar} | \psi_{\tau_M}} = \langle\psi^{(n+1)}_{tar}|\partial_{\phi_{n}}e^{-i[H(\tau_n,\phi_{n})]\Delta}|\psi^{(n-1)}_{init}\rangle.
\label{fidelderivative}
\end{align}
For each $n$, a way to calculate Eq.~(\ref{fidelderivative}) exactly 
is via full eigendecomposition of $H(\tau_n,\phi_n)$  \cite{Manches2011}, but 
the numerics scales very unfavourably with system size. Instead, we use Lanczos time evolution: For small $\Delta$, a split-operator method approximates the derivative in Eq.~(\ref{fidelderivative}):
\begin{align}\nonumber
&\partial_{\phi_{n}} e^{-i[H_{0}+V_{oc}(\phi_n)]\Delta}\approx \partial_{\phi_{n}} (e^{-iH_{0}\frac{\Delta}{2}} e^{-iV_{oc}(\phi_{n})\Delta} 
e^{-iH_{0}\frac{\Delta}{2}})\\
&= e^{-iH_{0}\frac{\Delta}{2}}[-i\Delta \partial_{\phi_{n}}V_{oc}(\phi_n)] e^{-iV_{oc}({\phi_n})\Delta} e^{-iH_{0}\frac{\Delta}{2}},
\end{align}
where $\partial_{\phi_{n}}V_{oc}$ can be placed unambiguously as it commutes with $e^{-iV_{oc}\Delta}$. This expression is then calculated with Lanczos time propagation and 
used in Eq.~(\ref{fidelderivative}).
After the evolution is performed and $\mathcal{F}$ and $\nabla_{\phi} \mathcal{F}$
are computed,
the results are entered in the optimisation cycle \cite{LBFGS}.
\begin{figure}[tbh]
  \centering
\includegraphics[width=0.47\textwidth]{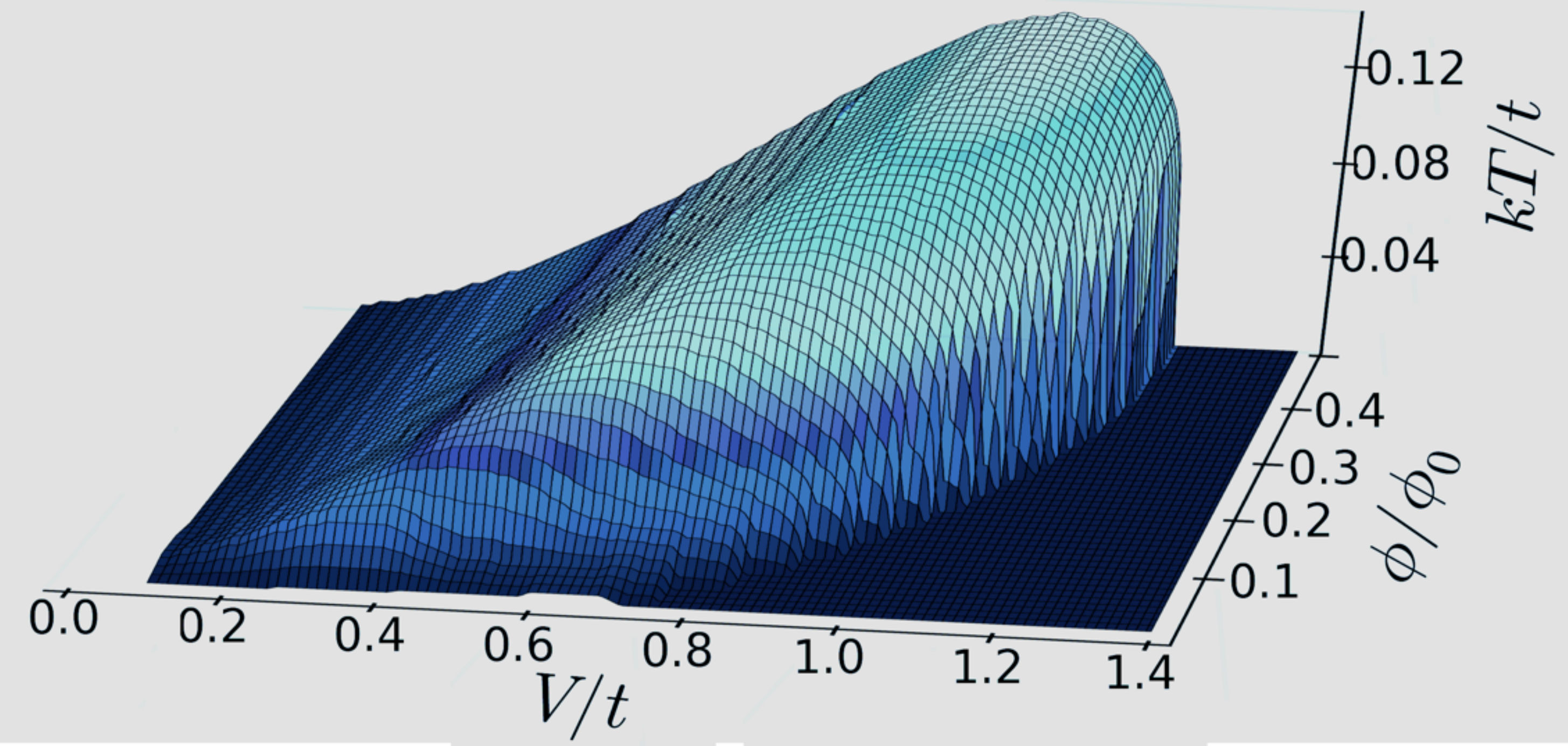}
\caption{Non-equilibrium Doniach-like phase diagram for a spin-compensated 4-site cluster as function of temperature $T$, Peierls phase $\phi$, hybridization parameter $V$.
The domain beneath the surface corresponds to the regime  with prevailing RKKY correlations.}
  \label{fig:dpd}
\end{figure}

{\bf Spin correlations vs steady currents at $T=0$. -}
In Fig.~\ref{fig:kondo}, we present results for spin-spin correlation functions (SCF) of $L=4$ and 6 homogeneous rings at zero temperature. The SCF considered are $\langle \hat{S}^z_c  \hat{S}^z_f \rangle$ (local-Kondo SCF) and $\langle \hat{S}^z_f  \hat{S}^z_{f'} \rangle$ (RKKY SCF), where $f,f'$ refer to impurities at nearest-neighbour sites in a ring. The groundstate of these systems is a singlet ($S=0$) and $2\pi/L$-periodic with respect to the Peierls phase $\phi$. For $L=4$, Kondo (RKKY) correlations are lowered (increased) when the Peierls phase $\phi$ approaches half of the flux quantum. For the $L=6$ ring we have an opposite trend, with the
roles of Kondo and RKKY inverted.

The opposite trends for $L=4$ and 6 are due to the way exchange interactions are affected by $\phi$. To clarify this, we consider an $L$-site conduction ring at half-filling isolated from the impurities, and denote by $\Lambda$ the difference between the highest occupied and lowest unoccupied energy levels in the ring. The
one-electron wave functions in the ring are $\psi_{k_m} \propto e^{ik_mR}$, with
$k_m = \frac{2\pi}{L}m$, and $m \in \left\lbrace 0,\ldots, L-1 \right\rbrace$. Adding a Peierls phase shifts the energy levels: $E_{k_m}(\phi)=-2\cos(\frac{2\pi}{L}(m+\phi/\phi_{0}))$ (Fig.~\ref{fig:kondo}, bottom panels). For $L=4$ and $\phi=0$, the lowest energy level is occupied by two electrons with opposite spin, and the next two levels are degenerate and half occupied, i.e. $\Lambda=0$. When $\phi/\phi_{0}\rightarrow 0.5$, the lowest two energy levels become degenerate and fully occupied by the four conduction electrons, i.e. $\Lambda>0$.
The half filled 6-site ring has opposite behavior: At $\phi=0$, the three lowest levels are doubly occupied, and $\Lambda=2t$. For $\phi/\phi_{0}=0.5$, the system reverts to partially filled degenerate levels and $\Lambda=0$. Going back to the interacting PAM ring,
the dependence of $\Lambda$ on $\phi$ affects the exchange interactions and thus the SCF: when the conduction 
electrons are locked in their levels ($\Lambda>0$), the spin exchange with an impurity atom is hindered, and so are in turn Kondo correlations ($\phi/\phi_{0}=0.5$ for the 4-site ring and $\phi/\phi_{0}=0$ for the 6-site ring). Conversely, when $\Lambda=0$ ($\phi/\phi_{0}=0$ for $L=4$ and $\phi/\phi_{0}=0.5$ for $L=6$), electrons participate more easily in spin-exchange processes, and this increases Kondo correlations.

For large $V$ the current dies out, due to robust
Kondo singlets, whilst at small $V$ the current tends to
that of a non interacting ring without impurities.
For different values of $\phi$ the current dependence on $V$ changes:  for example,
for $L=4$, the current drops much faster at lower flux (not shown), in line with the faster crossing in Fig.~\ref{fig:kondo} from RKKY to Kondo for $\phi/\phi_0=0.2$. 

{\bf $T>0$: non-equilibrium Doniach-like phase diagram.-} We now address the joint role of thermal fluctuations and currents
on competing magnetic phases. Fig. \ref{fig:dpd} displays the generalization of a Doniach-like phase diagram \cite{Verdozzi2004,Luo2005,Samuelsson2007,Potthoff1,Potthoff2} to non-equilibrium
for an $L=4$ ring. \footnote{ \revision{\color{black}Steady ring-currents from constant Peierls
phases (and the related Doniach-like phase diagrams), can be computed via 
ground-state (or thermal) equilibrium calculations. However, it should also be considered how such currents are established. Here, we use either adiabatic or fast ramping of the Peierls phases, starting from a zero-current ground state. Accordingly, for simplicity, and for both types of ramping, we generically use the expression
"non-equilibrium" Doniach phase diagram, to also implicitly convey the notion that magnetic phases with and without currents can differ.}}%
In a macroscopic sample, for the phase-diagram one would consider the relation between
Kondo $T_K \propto \exp(-\frac{1}{n(E_F)J})$ and RKKY $T_{RKKY}\propto J^2 n(E_F)$ 
temperatures, with $n(E_F)$ the density of states of the conduction electrons at the Fermi energy, and $J$ the effective impurity-conduction electron coupling \cite{Doniach77}.
Here, as in previous work \cite{MeanField,Verdozzi2004,Luo2005,Samuelsson2007}, the phase-diagram surface is the boundary where Kondo and RKKY SCF have equal value. The RKKY regime corresponds to the region beneath the surface, with antiferromagnetically ordered impurities. 

Fig.~\ref{fig:dpd} shows that, depending on the parameters, the RKKY regime can be marginal when there is no flux, but greatly enhanced by varying the current (e.g. $\phi/\phi_0 = 0.5$ in Fig.~\ref{fig:dpd}). As for $T=0$, this is due to a large gap $\Lambda$ between occupied and unoccupied conduction energy levels: a much higher temperature is needed to excite the electrons and destroy the RKKY regime. 
Other features from the zero-flux case can persist in the presence of currents: for example, 
when $V< V_c$, i.e.  below a critical value, and $T$ is above a "non-equilibrium crossover" temperature $T^{ne.cr.}_{RKKY}$ (defined as the temperature where, for a given $\phi/\phi_0\neq 0$, the value of RKKY and Kondo SCF is the same in the ring), we have a domain of
free, disordered local moments. At the same time, for $V > V_c$, a cluster correspondent of a Kondo spin-liquid phase is attained, with independent local spin-singlets \revision{\color{black} \cite{localKondo}} along the ring. Here, a main effect of the current is moving such domain boundary $V_c(\phi)$, i.e. a 
nonequilibrium Doniach-like phase diagram shows how to tune the Peierls phase and move across regimes. 

\begin{figure}[t!]
\centering
\includegraphics[width=0.47\textwidth]{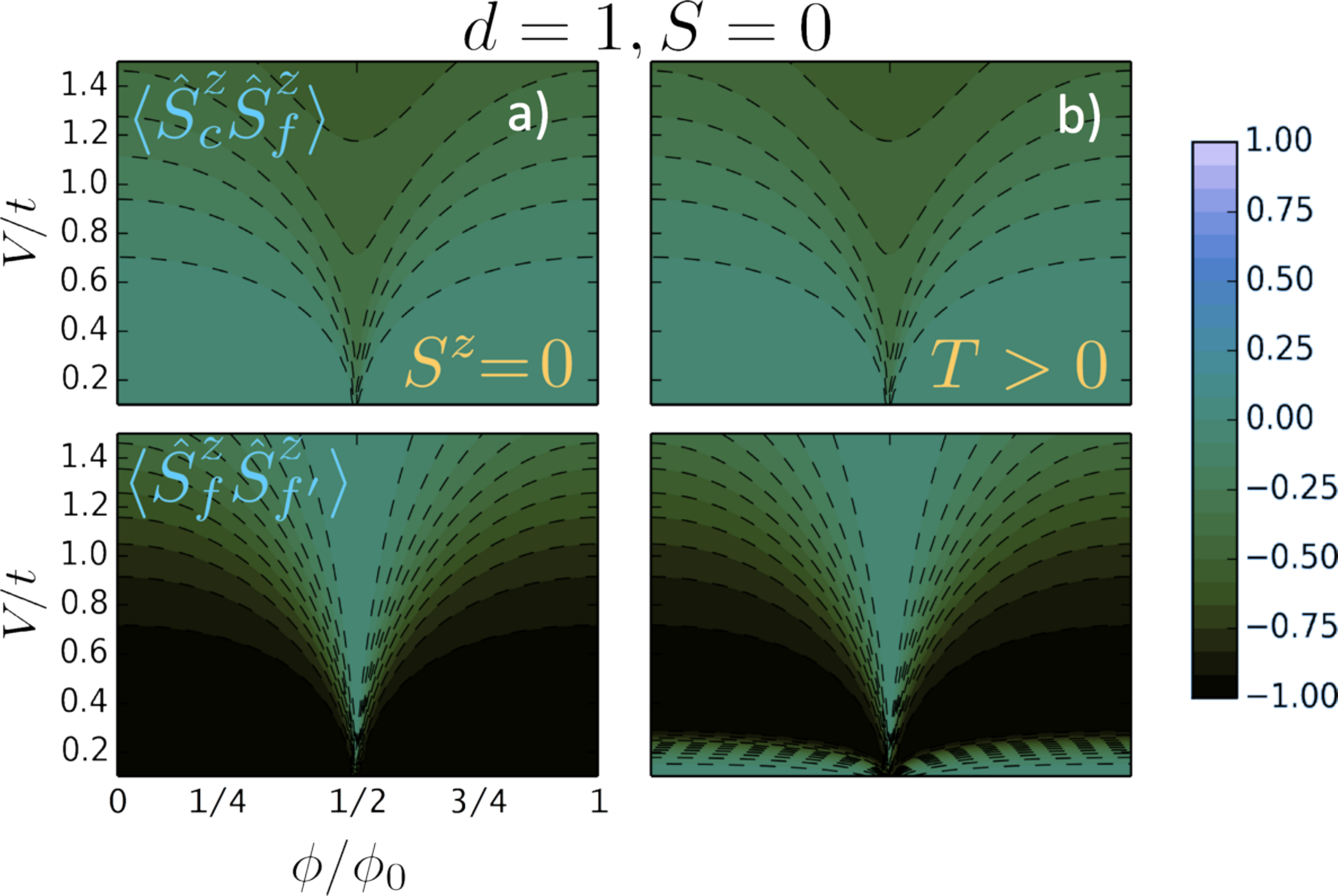}
  \caption{\revision{\color{black}Spin-spin correlation functions (SCF) for two impurities, $L=6$ and $d=1$ (cluster shown
in Fig.~\ref{fig:kondo}b). Panels in column a): $S=0,S^z=0$ ground state SCF; panels in column b): thermal averages. Panels in the the upper (lower) row pertain to Kondo (RKKY) SCF.
The scale for the Peierls phase $\phi/\phi_0$ in a) applies to all panels. The vertical bar on the right gives the value of the SCF. The dashed lines in the panels are a guide to the eye.
Thermal averages are for temperature $T_s=10^{-4}t/k_B$.}}
\label{fig:twoimp}
\end{figure} 

\begin{figure}[!b]
\centering
\includegraphics[width=0.47\textwidth]{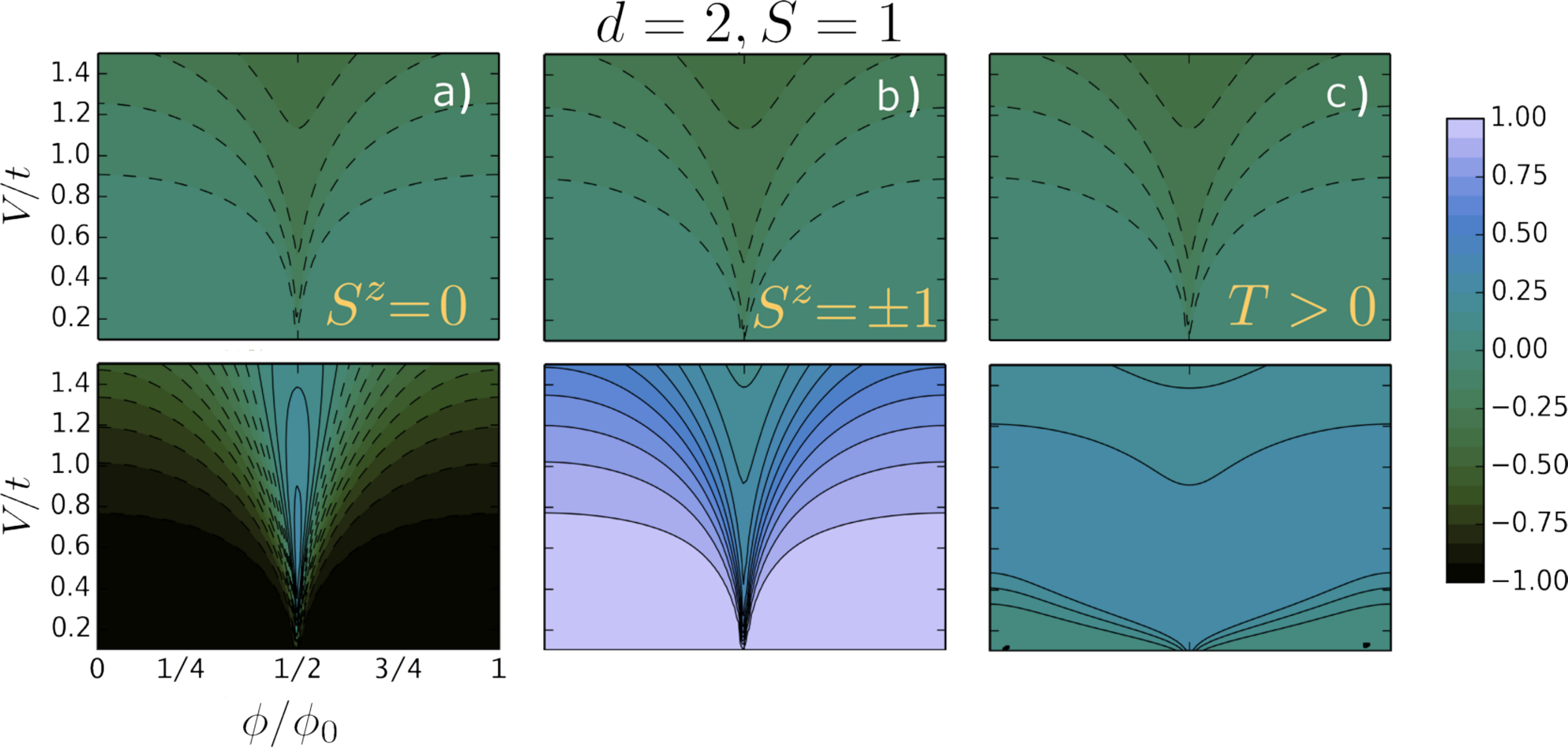}
  \caption{\revision{\color{black}Spin-spin correlation functions (SCF) for two impurities, $L=6$ and $d=2$ (cluster shown
  in Fig.~\ref{fig:kondo}c). Panels in columns a-b): $S=1,S^z=0,\pm 1$ ground state SCF; panels in column c): thermal averages. All other specifications are as in Fig.~\ref{fig:twoimp}.}} 
\label{fig:twoimp2}
\end{figure} 

Until now, the distance $d$ between impurities was fixed. \revision{\color{black}To asses the role of $d$, we consider a 6-site ring with two magnetic impurities, at half-filling. The situation 
with two nearest-neighbour impurities  ($d=1$, cluster shown in Fig.~\ref{fig:kondo}b) is reported in 
Figs.~\ref{fig:twoimp}, while the case of next-nearest-neighbour impurities ($d=2$, as shown in Fig.~\ref{fig:kondo}c) is illustrated in Fig.~\ref{fig:twoimp2}.} 

For $d=1$, the ground state is a singlet ($S=0$). \revision{\color{black} For thermal averages, we consider a temperature $k_BT_s/t=10^{-4}$ (this, for 
an hopping term of $t=1 \div 10$ eV, gives an actual value $T_s\approx 1\div10$ K). For such small $T_s$,
ground state and thermal SCF are very similar (panels in the left and right columns of Figs.~\ref{fig:twoimp}). This is quite expected,
since in this case the systems exhibits a finite, large gap between 
ground and first excited energy levels, and a very small temperature virtually plays no role.
We also note that $d=1$ we always observe negative Kondo and RKKY correlations.

The situation is rather different for $d=2$ (Fig.~\ref{fig:twoimp2}): here, due a $S=1$ (triplet) degeneracy in the ground state, it is appropriate to perform ensemble averages in the limit of zero temperature (hence
the small numerical value chosen for $T_s$). It is also worth mentioning, that for $d=2$ the RKKY thermal 
SCF are instead positive. 
 
Finally, from comparing Fig.~\ref{fig:twoimp} and Fig.~\ref{fig:twoimp2}, we note that the SCF strongly depend on the current (via $\phi$), the distance $d$, and the hybridization $V$. The dilute-impurity results just discussed
are for fixed cluster length $L$; concerning how results depend on $L$, an explorative set of selected calculations (not shown) suggests that, for a given distance $d$, there is an alternation of singlet-triplet ground states when $L$  is varied.}

{\bf Manipulating in time the exchange interaction via optimally controlled currents.-}
A nonequilibrium Doniach phase diagram gives the possibility of manipulating in time
exchange interactions, crossing from RKKY to Kondo regimes (and vice versa).
More in general, high proficiency in this crossing could be relevant to engineer new functionalities at the single- or few-spin level, as for example in memory switches. It is then quite natural to consider
optimal protocols to manipulate a system in time. Here, the more easily controllable parameter is the Peierls phase $\phi$, i.e. in practice the current (this has also the advantage of being accurately 
tunable in quantum transport setups).  

\begin{figure*}[tbh]
\centering
\includegraphics[width=1.0\textwidth]{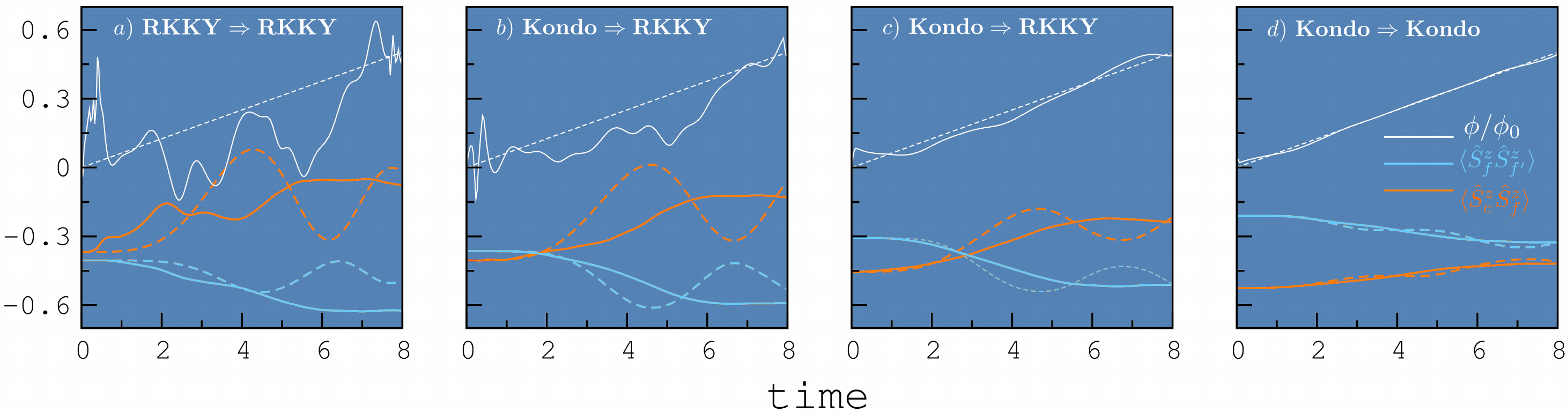}
  \caption{Kondo (orange curves) vs RKKY (light blue curves) correlations in a spin-compensated 4-site ring subject to a time-dependent flux/current (white curves) for a linear (dashed lines) and optimally controlled (solid lines) time protocol. The type of manipulation of the exchange interaction is indicated. 
In panels a-d), the values $(V/t, \mathcal{F})$
for the hybridization $V$ and fidelity $\mathcal{F}$ respectively are $(0.6,0.984)$, $(0.8,0.9999)$, $(1.0,0.9996)$, and $(1.3,0.99988$). Time is in units of the inverse hopping parameter.}
\label{fig:opt_con}
\end{figure*}

To switch from one regime to another, $\phi/\phi_0$ can either be changed smoothly (in principle infinitely slowly, i.e. adiabatically), or one can use optimal control theory \cite{Gross} and devise a current pulse so that target state is reached with maximal probability at the end of a chosen time interval $\tau_{M}$. To illustrate how this switching between regimes occurs in practice, we consider a 4-site ring at $T=0$ and compare
optimal and linear-ramping protocols in Fig.~\ref{fig:opt_con}. To illustrate different types of crossing/manipulation of the exchange interaction, each of the panels a-d) corresponds to a different value of the hybridization $V/t$ in the $(\phi\!-\!V)$ heatmap of Fig.~\ref{fig:kondo} when $L=4$. With $V/t$ fixed in a panel, the  $|\psi_{i}\rangle$ and $|\psi_{tar}\rangle$ of Eq.~(\ref{overl1},\ref{fidelderivative}) for the same panel are chosen as the singlet ground states corresponding to $\phi/\phi_0=0$ and $\phi/\phi_0=0.5$, respectively. Then the system is time-evolved until time $\tau_{M}$.

In Fig.~\ref{fig:opt_con},  we see that optimal pulses can have highly nonlinear temporal profiles, but there is stable saturation of the SCF at final time $\tau_{M}=8$. In contrast, for linear pulses there are induced (and not easily predictable in advance) oscillations in the SCF while the ramping is applied, and in general the desired state is not even attained at $\tau_{M}$. This confirms the importance and feasibility of optimal-control-based protocols to manipulate exchange via currents in a pre-established manner. 

The fidelity function $\mathcal{F}$ at $\phi/\phi_0= 0.5$ is essentially 1 in panels b-d), but $\lesssim 1$ in  a).
This suggests that when starting from the RKKY regime ($V/t=0.6$), it is much harder
to drive the system to a strong RKKY antiferromagnetic target state: the
process thus requires stronger (and marginally less successful) manipulation, as evident from the
highly non monotonic profile of $\phi(\tau)/\phi_0$. The connection between target exchange regime
and optimal flux/current is also apparent in panels b) and c): we see that reaching stronger (weaker) Kondo correlations at $\phi/\phi_0=0.5$, requires a weaker (stronger) manipulation.

The case $RKKY\Rightarrow Kondo$ is absent from Fig.~\ref{fig:opt_con}: in the diagram for $\langle \hat{S}^z_f  \hat{S}^z_{f'} \rangle$ and $L=4$, when increasing $\phi/\phi_0$ along a horizontal line (i.e. at constant $V/t$), the evolution $RKKY\Rightarrow Kondo$ is not possible. On the other hand, such crossing can occur for a $L=6$ ring ( Fig.~\ref{fig:kondo}). Because of size effects, it is plausible that the type of exchange switching which is possible to perform is quite dependent on the nature of the levels and size/shape of the nanosystem at hand.

{\bf Conclusions and Outlook.-}
In this work we showed how it is possible to dynamically manipulate
the nonequilibrium interplay of Kondo and RKKY exchange for systems with dense magnetic impurities in metallic hosts. To illustrate our point, we used charge current induced by magnetic fluxes in nanorings,
where currents alter the energy gap between occupied and empty levels and thus the magnetic interactions.
We used exact diagonalization, so we considered rather small systems. These however seem to be enough to illustrate the plausibility of our proposal, and generic aspects of the dynamical (and optimally controlled) manipulation of exchange at the nanoscale. It would be of interest to extend our work and e.g. consider the role of disorder, the behavior of entanglement, how optimal control of exchange relates to entropy production, and geometries of more direct technological relevance (such as dense impurities in quantum transport setups; in that case, on speculative grounds, we expect that tuning a nonequilbrium Kondo-vs-RKKY behavior should still be attainable, possibly with a different type of manipulation). It would also be
interesting to drive by currents a system to a point in the nonequilibrium Doniach phase diagram, and see the response to a current quench. 
\revision{\color{black} At this stage, we are aware of no actual experiments that have directly addressed RKKY vs Kondo 
physics and Doniach-like phase diagrams in nanorings in the presence of electric currents.  On purely speculative grounds,  potential experimental setups could be realised with optical lattices and artificial gauge fields \revision{\color{black} \cite{Cappellini}}.
Another possibility could be offered by a phenomenon called current magnification (used e.g.
to create persistent currents in graphene nanorings) \revision{\color{black} \cite{Benjamin}}.
Alternatively, quantum-dot transport setups used for studying the interference between Kondo tunneling and the Aharonov-Bohm effect might also be considered \revision{\color{black} \cite{Kikoin}}.}
Hopefully, our work will provide hints to other theoretical and experimental studies,
exploring further the Kondo-RKKY crossover and its possible relevance for technological applications.
\\
\begin{acknowledgments}
\end{acknowledgments}


\begin{thebibliography}{[10]}
\bibitem{localKondo} For the dense impurity case, expressions such as
"local Kondo singlet" are commonly used in the literature, and often for brevity the adjective ``local'' can even omitted. Here we follow the same practice.
\end{thebibliography}
\end{document}